\documentclass[a4paper,12pt]{article}

\usepackage{a4}
\usepackage{latexsym, amssymb}

\usepackage{graphicx}
\usepackage{epsfig}

\newcommand{\be}{\begin{equation}}
\newcommand{\ee}{\end{equation}}
\newcommand{\bea}{\begin{eqnarray}}
\newcommand{\eea}{\end{eqnarray}}
\newcommand{\mev}{\mbox{MeV}}
\newcommand{\gev}{\mbox{GeV}}
\newcommand{\VVP}{\langle V V \! P\rangle}
\def\order#1{{\cal O}\left(#1\right)}

\def\thefootnote{\fnsymbol{footnote}}

\begin{document}

\begin{titlepage}

\begin{flushright}
{\small 
April 28, 2009 \\
HRI-P-09-01-001 \\
RECAPP-HRI-2009-001}
\end{flushright}

\vspace*{0.2cm}
\begin{center}
{\Large {\bf 
Hadronic light-by-light scattering in the muon $g-2$: \\[0.2cm]
a new short-distance constraint on pion exchange}} \\[10mm]
Andreas Nyf\/feler\footnote{nyf\/feler@hri.res.in} \\[0.5cm]

{\it Regional Centre for Accelerator-based Particle Physics \\
Harish-Chandra Research Institute \\
Chhatnag Road, Jhusi \\ 
Allahabad - 211 019, India} \\[20mm]
\end{center}

\begin{abstract}
Recently it was pointed out that for the evaluation of the numerically
dominant pion-exchange contribution to the hadronic light-by-light scattering
correction in the muon $g-2$, a fully off-shell pion-photon-photon form factor
should be used. Following this proposal, we first derive a new short-distance
constraint on the off-shell form factor which enters at the external vertex
for the muon $g-2$ and show that it is related to the quark condensate
magnetic susceptibility in QCD. We then evaluate the pion-exchange
contribution in the framework of large-$N_C$ QCD using an off-shell form
factor which fulfills all short-distance constraints. With a value for the
magnetic susceptibility as estimated in the same large-$N_C$ framework, we
obtain the result $a_{\mu}^{\mathrm{LbyL};\pi^0} = (72 \pm 12) \times
10^{-11}$. Updating our earlier results for the contributions from the
exchanges of the $\eta$ and $\eta^\prime$ using simple vector-meson dominance
form factors, we obtain $a_{\mu}^{\mathrm{LbyL; PS}} = (99 \pm 16) \times
10^{-11}$ for the sum of all light pseudoscalars. Combined with available
evaluations for the other contributions to hadronic light-by-light scattering
this leads to the new estimate $a_{\mu}^{\mathrm{LbyL; had}} = (116 \pm 40)
\times 10^{-11}$.
\end{abstract}

\pagestyle{plain}

\end{titlepage}

\setcounter{page}{1}


\renewcommand{\thefootnote}{\arabic{footnote}}
\setcounter{footnote}{0}

\section{Introduction}
\label{sec:intro}

The hadronic light-by-light scattering contribution to the muon $g-2$ has a
long and troubled history. The relevant physics involves the nonperturbative
regime of QCD below about $1-2$~GeV. Furthermore, no direct experimental
information is available, in contrast to the hadronic vacuum polarization
contribution to the $g-2$, which can be related to the cross section $e^+ e^-
\to~\mbox{hadrons}$. Therefore various models have been used over the years,
starting first with a simple loop of some constituent
quarks~\cite{Calmet76}. Later more realistic hadronic models with exchanges of
light pseudoscalars, scalars and other resonances and loops with charged pions
have been employed~\cite{KNO85}. However, the coupling of hadrons to photons
will, in general, involve some form factors which are very model-dependent
[$\rho-\gamma$ mixing as in vector-meson dominance (VMD) models]. In the
absence of any direct experimental checks, the size and even the sign of the
light-by-light scattering contribution to the muon $g-2$ was therefore
uncertain for a long time. Actually, the sign has changed several times over
the years due to some errors in the complicated calculations.

In Ref.~\cite{EdeR94} a systematic approach was proposed, based on the chiral
expansion~\cite{ChPT} and the large-$N_C$ counting~\cite{largeNc} of the
various contributing diagrams. Soon afterwards, two very extensive evaluations
appeared, Refs.~\cite{BPP95,HKS95,HK98}, based on slightly different hadronic
models. However, they both had a sign error in the numerically dominating
pseudoscalar-exchange contribution as was pointed out a few years later in
Refs.~\cite{KN01,KNPdeR01} and confirmed in
Refs.~\cite{Bardeen_deGouvea,BCM01,RMW02}.

Reference~\cite{KN01} mainly concentrated on the neutral pion-exchange
contribution where the pion-photon-photon form factor ${\cal
F}_{{\pi^0}^*\gamma^*\gamma^*}$ enters. In general, low-energy hadronic models
for form factors, e.g.\ based on some constituent quark model or on some
resonance Lagrangian, do not satisfy all the large momentum asymptotics
required by QCD. Using these form factors in loop diagrams thus leads to
cutoff-dependent results. Even if the cutoff is varied in a reasonable range,
e.g.\ $\sim 1-2~\mbox{GeV}$, the corresponding model uncertainty is completely
uncontrollable.  In order to eliminate (or at least reduce) this cutoff
dependence, new models for ${\cal F}_{{\pi^0}^*\gamma^*\gamma^*}$ were
proposed in Ref.~\cite{KN_EPJC01} and then applied to hadronic light-by-light
scattering in Ref.~\cite{KN01}.  These models are based on the large-$N_C$
picture of QCD~\cite{largeNc}, where, in leading order in $N_C$, an (infinite)
tower of narrow resonances contributes in each channel of a particular Green's
function. The low-energy and short-distance behavior of these Green's
functions is then matched with results from QCD, using chiral perturbation
theory~\cite{ChPT} and the operator product expansion (OPE)~\cite{OPE},
respectively. Based on the experience gained in many examples of low-energy
hadronic physics, and from the use of dispersion relations and spectral
representations for two-point functions, it is then assumed that with a
minimal number of resonances in a given channel one can get a reasonable good
description of the QCD Green's function in the real world (minimal hadronic
Ansatz). Often only the lowest-lying resonance is considered [lowest-meson
dominance (LMD)]~\cite{MoussallamStern,Moussallam95,Moussallam97,LMD1,LMD2},
as a generalization of VMD.

Reference~\cite{KN01} obtained the result $a_{\mu}^{\mathrm{LbyL};\pi^0} = (58
\pm 10) \times 10^{-11}$ for the pion and $a_{\mu}^{\mathrm{LbyL;PS}} = (83
\pm 12) \times 10^{-11}$ for the sum of all light pseudoscalars $\pi^0, \eta$,
and $\eta^\prime$. These results are close to the (sign corrected) values
$a_{\mu}^{\mathrm{LbyL};\pi^0} = (59 \pm 9) \times 10^{-11}$
    [$a_{\mu}^{\mathrm{LbyL; PS}} = (85 \pm 13) \times 10^{-11}$] obtained in
    Ref.~\cite{BPP95} and $a_{\mu}^{\mathrm{LbyL};\pi^0} = (57 \pm 4) \times
    10^{-11}$ [$a_{\mu}^{\mathrm{LbyL; PS}} = (82.7 \pm 6.4) \times 10^{-11}$]
    in Refs.~\cite{HKS95,HK98}. The results for the (corrected) full
    contributions at that time read $a_{\mu}^{\mathrm{LbyL; had}} = (83 \pm
    32) \times 10^{-11}$~\cite{BPP95} and $a_{\mu}^{\mathrm{LbyL; had}} =
    (89.6 \pm 15.4) \times 10^{-11}$~\cite{HKS95,HK98}.

Later Ref.~\cite{MV03} pointed out that some additional QCD short-distance
constraint was not taken into account for the exchanges of pseudoscalars and
axial-vector resonances in Refs.~\cite{BPP95,HKS95,HK98,KN01}. The authors of
Ref.~\cite{MV03} argued that if one imposes this constraint, no
momentum-dependent form factor can be present at the external vertex which
couples to the soft photon relevant for the magnetic moment.  In the absence
of such a form factor, Ref.~\cite{MV03} then got enhanced results compared to
the earlier evaluations $a_{\mu}^{\mathrm{LbyL};\pi^0} = (77 \pm 7) \times
10^{-11}$ [$a_{\mu}^{\mathrm{LbyL; PS}} = (114 \pm 10) \times 10^{-11}$] and
$a_{\mu}^{\mathrm{LbyL; had}} = (136 \pm 25) \times 10^{-11}$. As discussed in
Ref.~\cite{BP07}, a part of the enhancement of the result for $a_\mu^{\rm
LbyL; had}$ in Ref.~\cite{MV03} is actually due to a different treatment of
the axial vectors (ideal mixing instead of nonet symmetry) and the omission of
the negative contributions from scalar exchanges and the charged pion loop. It
is therefore not entirely related to the new short-distance constraint. Thus
the slightly lower estimate $a_{\mu}^{\mathrm{LbyL; had}} = (110 \pm 40)
\times 10^{-11}$ has been employed in the reviews~\cite{BP07,MdeRR07}. Very
recently, the value $a_{\mu}^{\mathrm{LbyL; had}} = (105 \pm 26) \times
10^{-11}$ has been proposed in Ref.~\cite{Prades_deRafael_Vainshtein09}.

However, recently Refs.~\cite{FJ_Essentials, FJ_Book} stressed the fact that
one should actually use fully off-shell form factors for the evaluation of the
light-by-light scattering contribution. This seems to have been overlooked in
the recent literature, in particular, in
Refs.~\cite{KN01,Bijnens_Persson01,MV03,BP07}. The on-shell form factors as
used in Refs.~\cite{KN01,Bijnens_Persson01} actually violate four-momentum
conservation at the external vertex. While Ref.~\cite{MV03} had already
pointed out this violation of momentum conservation at the external vertex,
they then only considered {\it on-shell} pion form factors, an approximation
which yields the so-called {\it pion-pole} contribution and not the more
general {\it pion-exchange} contribution with off-shell form factors. Putting
the pion on-shell at the external vertex automatically leads to a constant
form factor.

In the present paper we revisit the pion-exchange contribution in view of the
observations made in Refs.~\cite{FJ_Essentials, FJ_Book}.  We first derive a
new QCD short-distance constraint for the off-shell form factor which enters
at the external vertex and show that it is related to the quark condensate
magnetic susceptibility in QCD. We also comment on how our short-distance
constraint is connected with the one derived in Ref.~\cite{MV03}. In the
second part we evaluate the pion-exchange contribution in the framework of
large-$N_C$ QCD with off-shell form factors both at the internal and the
external vertex, taking into account the new short-distance constraint and an
estimate for the magnetic susceptibility in QCD in the same large-$N_C$
framework.

Strictly speaking, the identification of the pion-exchange contribution is
only possible, if the pion is on-shell (or nearly on-shell). If one is (far)
off the mass shell of the exchanged particle, it is not possible to separate
different contributions to the $g-2$, unless one uses some particular model
where for instance elementary pions can propagate. In this sense, only the
pion-pole contribution with on-shell form factors can be defined, at least in
principle, in a model-independent way, although the numerical result will in
general still depend on the model used for the on-shell form factors, unless
one would know the ``true'' form factors. On the other hand, the pion-pole
contribution is only a part of the full result, since in general, e.g.\ using
some resonance Lagrangian, the form factors will enter the calculation with
off-shell momenta. In this respect, we view our evaluation as being a part of
a full calculation of hadronic light-by-light scattering using a resonance
Lagrangian whose coefficients are tuned in such a way as to systematically
reproduce the relevant QCD short-distance constraints, e.g.\ along the lines
of the resonance chiral theory developed in Ref.~\cite{Ecker_etal}.

We should mention that recently another paper
appeared~\cite{Dorokhov_Broniowski} which evaluates the pion-exchange
contribution using an off-shell form factor based on the nonlocal chiral quark
model, obtaining the result $a_{\mu}^{\mathrm{LbyL};\pi^0} = (65 \pm 2) \times
10^{-11}$. We will comment on that paper below.

This paper is organized as follows. Section~\ref{sec:pionpole} contains the
starting point for the calculation of the pion-exchange contribution to the
muon $g-2$, including the definition of the pion-photon-photon form factor
${\cal F}_{{\pi^0}^*\gamma^*\gamma^*}$. We also discuss the issue of using
on-shell or off-shell form factors. In Sec.~\ref{sec:FF_constraints} we
discuss several experimental and theoretical constraints on the form
factor. In particular, we derive a new short-distance constraint on the
off-shell form factor at the external vertex in hadronic light-by-light
scattering. In Sec.~\ref{sec:new_evaluation_pseudoscalars} we present a new
evaluation of the pion-exchange contribution in the framework of large-$N_C$
QCD and give some updated estimates for the $\eta$ and $\eta^\prime$ exchange
contributions using simple VMD form factors. We end with discussions and
conclusions in Sec.~\ref{sec:conclusions}. In the appendix we give a
parametrization of the numerical result for the pion-exchange contribution for
arbitrary parameters of our model for the off-shell form factor.

\section{The pseudoscalar-exchange contribution}
\label{sec:pionpole}

\begin{figure}[h]
\centering
\centerline{\epsfig{figure=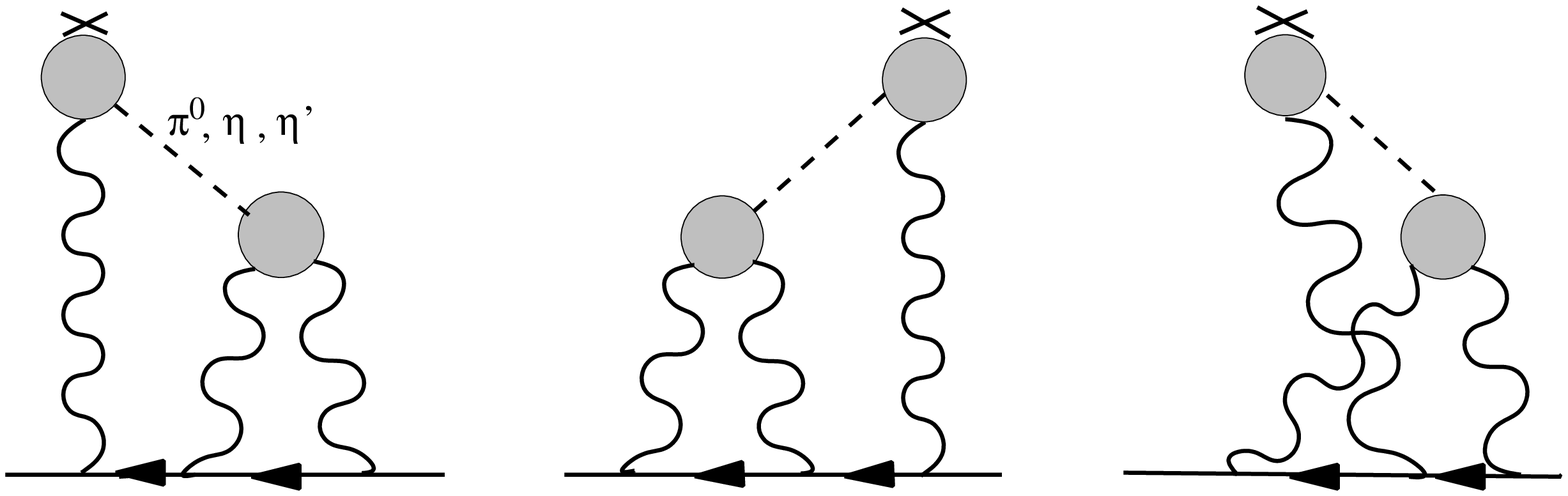,height=4.5cm,width=15cm}}
\caption{The pseudoscalar-exchange contributions to hadronic light-by-light
  scattering. The shaded blobs represent the off-shell form factor ${\cal
  F}_{{\rm PS}^* \gamma^* \gamma^*}$ where ${\rm PS} = \pi^0, \eta,
  \eta^\prime$.}
\label{fig:LbL_PS-exchanges}
\end{figure}

The numerically dominating contributions to hadronic light-by-light
scattering are due to the neutral pseudoscalar-exchange diagrams shown in
Fig.~\ref{fig:LbL_PS-exchanges}. 

We first concentrate on the exchange of the neutral pion.  The key object
which enters the Feynman diagrams in Fig.~\ref{fig:LbL_PS-exchanges} is the
{\it off-shell} pion-photon-photon form factor ${\cal
F}_{{\pi^0}^*\gamma^*\gamma^*}((q_1+q_2)^2,q_1^2, q_2^2)$ which is defined, up
to small mixing effects with the states $\eta$ and $\eta^\prime$, via the
Green's function $\VVP$ in QCD
\bea
\lefteqn{ \int d^4 x\, d^4 y \, e^{i (q_1 \cdot x + q_2 \cdot y)} \, 
\langle\,0 | T \{ j_\mu(x) j_\nu(y) P^3(0) \} | 0 \rangle \, } \nonumber \\  
&=&  \varepsilon_{\mu\nu\alpha\beta} \, q_1^\alpha q_2^\beta \, 
 {i \langle{\overline\psi}\psi\rangle \over F_\pi} \, {i \over (q_1 +
   q_2)^2 - m_\pi^2} \, {\cal F}_{{\pi^0}^*\gamma^*\gamma^*}((q_1 +
 q_2)^2,q_1^2,q_2^2).  
\label{FFoffshellpi}
\eea
Here $j_\mu(x) = ({\overline \psi} \hat Q \gamma_\mu \psi)(x)$ [${\overline
  \psi} \equiv ({\overline u}, {\overline d}, {\overline s}$), $\hat Q =
  \mbox{diag}(2,-1,-1)/3$ the charge matrix] is the light quark part of the
  electromagnetic current and $P^3 = {\overline \psi} i \gamma_5 {\lambda^3
  \over 2} \psi = \left( {\overline u} i\gamma_5 u - {\overline d} i \gamma_5
  d \right) / 2$. Note that we denote by $\langle{\overline\psi}\psi\rangle$
  the {\it single flavor} bilinear quark condensate. The form factor is of
  course Bose symmetric ${\cal
  F}_{{\pi^0}^*\gamma^*\gamma^*}((q_1+q_2)^2,q_1^2,q_2^2) = {\cal
  F}_{{\pi^0}^*\gamma^*\gamma^*}((q_1+q_2)^2,q_2^2,q_1^2)$, as the two photons
  are indistinguishable.

The corresponding contribution to the muon $g-2$ may be worked out with the
result~\cite{KN01}\footnote{To be precise, the corresponding
  expression with {\it on-shell} form factors is given in
  Ref.~\cite{KN01}.}  
\bea
a_{\mu}^{\mathrm{LbyL};\pi^0}& = & - e^6
\int {d^4 q_1 \over (2\pi)^4} {d^4 q_2 \over (2\pi)^4}
\,\frac{1}{q_1^2 q_2^2 (q_1 + q_2)^2[(p+ q_1)^2 - m_\mu^2][(p - q_2)^2 -
    m_\mu^2]} 
\nonumber \\
&& 
\times \left[
{{\cal F}_{{\pi^0}^*\gamma^*\gamma^*}(q_2^2, q_1^2, (q_1 + q_2)^2) \ {\cal 
    F}_{{\pi^0}^*\gamma^*\gamma}(q_2^2, q_2^2, 0) \over q_2^2 - 
m_{\pi}^2} \ T_1(q_1,q_2;p) \nonumber \right. \\
&& 
\  + \left. {{\cal F}_{{\pi^0}^*\gamma^*\gamma^*}((q_1+q_2)^2, q_1^2,  q_2^2)
  \ {\cal F}_{{\pi^0}^*\gamma^*\gamma}((q_1+q_2)^2, (q_1+q_2)^2, 0) \over
  (q_1+q_2)^2 - m_{\pi}^2} \ T_2(q_1,q_2;p) \right], \nonumber \\
& & 
\label{a_pion_2}
\eea
with
\bea
T_1(q_1,q_2;p) & = & {16 \over 3}\, (p \cdot q_1) \, (p \cdot q_2) \,
(q_1 \cdot q_2)
\,-\, {16 \over 3}\, (p \cdot q_2)^2 \, q_1^2 \nonumber \\
&& 
-\, {8 \over 3}\, (p \cdot q_1) \, (q_1 \cdot q_2) \, q_2^2
\,+\, 8 (p \cdot q_2) \, q_1^2 \, q_2^2
\,-\,{16 \over 3} (p \cdot q_2) \, (q_1 \cdot q_2)^2 \nonumber
\\
&& 
+\, {16 \over 3}\, m_\mu^2 \, q_1^2 \, q_2^2
\,-\, {16 \over 3}\, m_\mu^2 \, (q_1 \cdot q_2)^2 \, ,  \\
T_2(q_1,q_2;p) & = & {16 \over 3}\, (p \cdot q_1) \, (p \cdot q_2) \,
(q_1 \cdot q_2) \,-\,{16 \over 3}\, (p \cdot q_1)^2 \, q_2^2
\nonumber \\
&& 
 +\, {8 \over 3}\, (p \cdot q_1) \, (q_1 \cdot q_2) \, q_2^2
\,+\, {8 \over 3}\, (p \cdot q_1) \, q_1^2 \, q_2^2
\,\nonumber \\
&& 
 +\, {8 \over 3}\, m_\mu^2 \, q_1^2 \, q_2^2
\,-\, {8 \over 3}\, m_\mu^2 \, (q_1 \cdot q_2)^2 \, , 
\eea
where $p^2 = m_\mu^2$ and the external photon has now zero four-momentum. The
first and the second graphs in Fig.~\ref{fig:LbL_PS-exchanges} give rise to
identical contributions, leading to the term with $T_1$, whereas the third
graph gives the contribution involving $T_2$.  The factor $T_2$ has been
symmetrized with respect to the exchange $q_1\leftrightarrow -q_2$.

Instead of the expressions in Eq.~(\ref{a_pion_2}),
Refs.~\cite{KN01,Bijnens_Persson01} and maybe also earlier works,
considered on-shell form factors, e.g.\ for the term involving $T_2$ one would
write~\cite{FJ_Book}  
\be
{\cal F}_{\pi^0 \gamma^* \gamma^*}(m_\pi^2,
q_1^2,q_2^2) \ \times \ {\cal F}_{\pi^0 \gamma^* \gamma}(m_\pi^2,
(q_1 + q_2)^2,0). 
\ee
Often the first argument of the on-shell form factor is omitted in the
literature, i.e.\ the form factor is written as ${\cal F}_{\pi^0 \gamma^*
\gamma^*}(q_1^2,q_2^2) \equiv {\cal F}_{\pi^0 \gamma^* \gamma^*}(m_\pi^2,
q_1^2, q_2^2)$.  Although pole dominance might be expected to give a
reasonable approximation, it is not correct as it was used in those
references, as stressed in Refs.~\cite{MV03,FJ_Essentials,FJ_Book}.  The point
is that the form factor sitting at the external photon vertex in the pole
approximation ${\cal F}_{\pi^0 \gamma^* \gamma}(m_\pi^2,(q_1 + q_2)^2,0)$ for
$(q_1 + q_2)^2 \neq m_\pi^2$ violates four-momentum conservation, since the
momentum of the external (soft) photon vanishes.  The latter requires ${\cal
F}_{{\pi^0}^* \gamma^* \gamma}((q_1 + q_2)^2, (q_1 + q_2)^2, 0)$. In order to
avoid this inconsistency, Ref.~\cite{MV03} proposed to use instead
\be
{\cal F}_{\pi^0 \gamma^* \gamma^*}(m_\pi^2,
q_1^2,q_2^2) \ \times \ {\cal F}_{\pi^0 \gamma \gamma}(m_\pi^2,
m_\pi^2,0)\,, 
\ee
i.e.\ a constant form factor at the external vertex, which is given by the
Wess-Zumino-Witten (WZW) anomaly~\cite{WZW}. The absence of a form factor at
the external vertex in the pion-pole approximation follows automatically, if
one carefully considers the momentum dependence of the form factor. This
procedure is also consistent with any quantum field theoretical framework for
hadronic light-by-light scattering, for instance, if one uses a (resonance)
Lagrangian to derive the form factors, and where a different treatment of the
internal and external vertex, apart from the kinematics, is not possible. On
the other hand, taking the diagram more literally, would require
\be \label{FFxFF_offshell}
{\cal F}_{{\pi^0}^* \gamma^* \gamma^*}((q_1 + q_2)^2, q_1^2,q_2^2) \ \times \
{\cal F}_{{\pi^0}^* \gamma^* \gamma}((q_1 + q_2)^2, (q_1 + q_2)^2,0)\,, 
\ee
as the more appropriate amplitude, see Eq.~(\ref{a_pion_2}).
References~\cite{FJ_Essentials,FJ_Book} advocate the use of fully off-shell
form factors at both vertices and we will follow this approach in the rest of
this paper. The difference to the procedure adopted in Ref.~\cite{MV03} will
be important when we discuss their short-distance constraint.

\section{Experimental and theoretical constraints on the pion-photon-photon 
  form factor}
\label{sec:FF_constraints}

The form factor ${\cal
F}_{{\pi^0}^*\gamma^*\gamma^*}((q_1+q_2)^2,q_1^2,q_2^2)$ defined in
Eq.~(\ref{FFoffshellpi}) is determined by nonperturbative physics of QCD and
cannot (yet) be calculated from first principles. Therefore, various
hadronic models have been used in the literature, sometimes combined with
short-distance constraints from perturbative QCD at high momenta. At low
energies, the form factor is normalized by the decay amplitude, ${\cal
A}(\pi^0 \to \gamma\gamma) \equiv e^2 {\cal F}_{\pi^0\gamma\gamma}(m_\pi^2, 0,
0) $ in our conventions.  In the chiral limit, $m_q \to 0, q=u,d,s$, this
amplitude is fixed by the WZW anomaly
\be
{\cal A}^{(0)}(\pi^0 \to \gamma\gamma) = - {e^2 N_C \over 12 \pi^2 F_0} . 
\ee
For massive light quarks, this expression receives corrections. In particular,
the pion decay constant in the chiral limit, $F_0$, is replaced by its
physical counterpart $F_\pi = F_0 [1 + \order{m_q}]$: 
\be
{\cal A}(\pi^0 \to \gamma\gamma) = - {e^2 N_C \over 12 \pi^2 F_\pi} [1
  + \order{m_q}] .  
\ee 
It turns out that the additional $\order{m_q}$ corrections in this relation
are numerically small~\cite{Pseudoscalar_decays}, so that one may drop them to
a good approximation. The measured decay width $\Gamma(\pi^0 \to \gamma\gamma)
= (7.74 \pm 0.6)~\mbox{eV}$~\cite{PDG2008} is then well reproduced for $F_\pi =
92.4~\mev$. Therefore, all hadronic models for the form factor have to satisfy
the low-energy constraint
\be \label{FF_normalization_WZW} 
{\cal F}_{\pi^0\gamma\gamma}(m_\pi^2, 0, 0) = - {N_C \over 12 \pi^2 F_\pi}. 
\ee
Sometimes this normalizing value is used to define a constant ``WZW form
factor'' ${\cal F}^{\rm WZW}_{{\pi^0}^*\gamma^*\gamma^*}((q_1+q_2)^2, q_1^2,
q_2^2) \equiv - N_C / (12 \pi^2 F_\pi)$. This notion of a constant form factor
is, however, very misleading. For off-shell momenta away from the physical
point in Eq.~(\ref{FF_normalization_WZW}) the value of this WZW form factor
has no physical meaning. Recall that the WZW effective Lagrangian only yields
the first term in the low-energy and chiral expansion of the corresponding
$\VVP$ Green's function. 

For an on-shell pion, there is also experimental data available for one
on-shell and one off-shell photon, from the process $e^+ e^- \to e^+ e^-
\pi^0$. Several experiments~\cite{CELLO90,CLEO98} thereby fairly well confirm
the Brodsky-Lepage~\cite{LepageBrodsky80} behavior for large Euclidean
momentum
\be \label{Brodsky_Lepage} 
\lim\limits_{Q^2 \to \infty} \: {\cal F}_{\pi^0 \gamma^*
  \gamma}(m_\pi^2,-Q^2,0) \sim - \frac{2 
F_\pi}{Q^2} 
\ee
and any satisfactory model should reproduce this behavior.

Apart from these experimental constraints, any consistent hadronic model for
the off-shell form factor ${\cal F}_{{\pi^0}^*\gamma^*\gamma^*}((q_1 + q_2)^2,
q_1^2, q_2^2)$ should match at large momentum with short-distance constraints
from QCD that can be calculated using the OPE. In Ref.~\cite{KN_EPJC01} the
short-distance properties for the three-point function $\VVP$ in
Eq.~(\ref{FFoffshellpi}) in the chiral limit and assuming octet symmetry have
been worked out in detail (see also Refs.~\cite{Moussallam95,LMD2} for earlier
partial results). At least for the pion the chiral limit should be a not too
bad approximation\footnote{As pointed out in Ref.~\cite{Nyffeler_RADCOR_2002},
the integrals in Eq.~(\ref{a_pion_2}) are infrared safe for $m_\pi \to
0$. This can also be seen within the effective field theory approach to
light-by-light scattering proposed in Refs.~\cite{KNPdeR01,RMW02}.}; however,
for the $\eta$ and, in particular, for the non-Goldstone boson $\eta^\prime$
further analysis will be necessary.

It is important to notice that the Green's function $\VVP$ is an order
parameter of chiral symmetry. Therefore, it vanishes to all orders in
perturbative QCD in the chiral limit, so that the behavior at short distances
is smoother than expected from naive power counting arguments.  Two limits are
of interest. In the first case, the two momenta become simultaneously large,
which in position space describes the situation where the space-time arguments
of all three operators tend towards the same point at the same rate. To
leading order and up to corrections of order $\order{\alpha_s}$ one obtains
the following behavior for the form factor\footnote{In the chiral limit, the
relation between the off-shell form factor and the single invariant function
${\cal H}_V$ which appears in $\VVP$ is given by ${\cal
F}_{{\pi^0}^*\gamma^*\gamma^*}((q_1 + q_2)^2, q_1^2, q_2^2) = - (2/3) (F_0 /
\langle{\overline\psi}\psi\rangle_0) (q_1 + q_2)^2 {\cal H}_V(q_1^2, q_2^2,
(q_1 + q_2)^2)$; see Ref.~\cite{KN_EPJC01} for details.}:
\be
\lim_{\lambda \to \infty} {\cal F}_{{\pi^0}^*\gamma^*\gamma^*}((\lambda q_1 +
\lambda q_2)^2, (\lambda q_1)^2, (\lambda q_2)^2) =  {F_0 \over 3} \, {1 \over
  \lambda^2} {q_1^2 + q_2^2 + (q_1+q_2)^2 \over q_1^2 q_2^2}  \nonumber +
\order{{1\over \lambda^4}} \, . \label{FF_OPE_1} 
\ee

The second situation of interest corresponds to the case where the relative
distance between only two of the three operators in $\VVP$ becomes small.  It
so happens that the corresponding behaviors in momentum space involve, apart
from the correlator $\langle A P\rangle$ which, in the chiral limit, is
saturated by the single-pion intermediate state
\be
\int d^4 x e^{i p \cdot x}
\langle 0 \vert T \{ A_\mu^a(x) P^b(0) \} \vert 0 \rangle =
\delta^{ab} \langle{\overline\psi}\psi\rangle_0 \, {p_\mu \over p^2}  
\ee   
(we denote by $\langle{\overline\psi}\psi\rangle_0$ the single flavor bilinear
  quark condensate in the chiral limit), the two-point function $\langle
  VT\rangle$ of the vector current and the antisymmetric tensor density
\be \label{Pi_VT}
\delta^{ab}(\Pi_{\rm VT})_{\mu\rho\sigma}(p)\,=\, 
\int d^4x e^{i p \cdot x}
\langle 0 \vert T \{ V_\mu^a(x) 
({\overline\psi}\,\sigma_{\rho\sigma}\frac{\lambda^b}{2}\,\psi)(0)\}\vert
0\rangle \, ,  
\ee
with $\sigma_{\rho\sigma}={i\over 2}[\gamma_{\rho},\gamma_{\sigma}]$
(the similar correlator between the axial current and the tensor
density vanishes as a consequence of invariance under charge conjugation). 
Conservation of the vector current and invariance under parity then give
\be \label{Pi_VT_invariant_function}
(\Pi_{\rm VT})_{\mu\rho\sigma}(p)\,=
\,(p_{\rho}\eta_{\mu\sigma}-p_{\sigma}\eta_{\mu\rho})\,\Pi_{\rm VT}(p^2)
\,. 
\ee

When the space-time arguments of the two vector currents in $\VVP$ approach  
each other, the leading term in the OPE leads to the Green's function $\langle
A P\rangle$ and the short-distance behavior of the form factor reads
\be
\lim_{\lambda \to \infty} {\cal F}_{{\pi^0}^*\gamma^*\gamma^*}(q_2^2, (\lambda
q_1)^2, (q_2-\lambda q_1)^2) = {2 F_0 \over 3} {1 \over \lambda^2} {1 \over
  q_1^2} + \order{{1\over \lambda^3}} .   \label{FF_OPE_2}
\ee

Further important information on the on-shell pion form factor has been
obtained in Ref.~\cite{ShuVai82} based on higher-twist terms in the OPE and
worked out in~\cite{NSVVZ84}. In the chiral limit one obtains the behavior
\be
\lim_{\lambda \to \infty}
\frac{{\cal F}_{{\pi^0}\gamma^*\gamma^*}(0,(\lambda q_1)^2,(\lambda
  q_1)^2)}{{\cal F}_{{\pi^0}\gamma\gamma}(0,0,0)}= 
- \frac{8}{3}\pi^2 F_0^2\left\{\frac{1}{\lambda^2 q_1^2}+\frac{8}{9}
\frac{\delta^2}{\lambda^4 q_1^4}+ \order{{1\over \lambda^6}} \right\}, 
\label{NSVVZ}
\ee
where $\delta^2$ parametrizes the relevant higher-twist matrix element.  
The sum rule estimate performed in~\cite{NSVVZ84} yields the value $\delta^2
=(0.2 \pm 0.02)~\gev^2$. 

On the other hand, when the space-time argument of one of the vector currents
in $\VVP$ approaches the argument of the pseudoscalar density one obtains the 
relation~\cite{KN_EPJC01}   
\be
\lim_{\lambda \to \infty} {\cal F}_{{\pi^0}^*\gamma^*\gamma^*}((\lambda q_1 +
q_2)^2, (\lambda q_1)^2,q_2^2) = - {2 \over 3} {F_0 \over
  \langle{\overline\psi}\psi\rangle_0} \Pi_{\rm VT}(q_2^2) 
+ \order{{1\over \lambda}} \, . \label{FF_OPE_3} 
\ee
In particular, at the external vertex in light-by-light scattering in
Eq.~(\ref{a_pion_2}), the following limit is relevant  
\be
\lim_{\lambda \to \infty} {\cal F}_{{\pi^0}^*\gamma^*\gamma}((\lambda
q_1)^2, (\lambda q_1)^2,0) = - {2 \over 3} {F_0 \over
  \langle{\overline\psi}\psi\rangle_0} \Pi_{\rm VT}(0) 
+ \order{{1\over \lambda}} \, .
\label{FF_OPE_3_zeromomentum}  
\ee
Note that there is no falloff in this limit, unless $\Pi_{\rm VT}(0)$
vanishes. 

As pointed out in Ref.~\cite{Belyaev_Kogan}, the value of $\Pi_{\rm VT}(p^2)$
  at zero momentum is related to the quark condensate magnetic susceptibility
  $\chi$ in QCD in the presence of a constant external electromagnetic field,
  introduced in Ref.~\cite{Ioffe_Smilga}: 
\be
\langle 0 | \bar{q} \sigma_{\mu\nu} q | 0 \rangle_{F} = e \, e_q \, \chi
\, \langle{\overline\psi}\psi\rangle_0 \, F_{\mu\nu}, 
\ee
with $e_u = 2/3$ and $e_d = -1/3$. With our definition of
$\Pi_{\rm VT}$ in Eq.~(\ref{Pi_VT}) one then obtains the relation (see also
Ref.~\cite{Mateu_Portoles})  
\be \label{Pi_VT0_Chi}
\Pi_{\rm VT}(0) = - {\langle{\overline\psi}\psi\rangle_0 \over 2} \chi, 
\ee
and therefore the behavior at the external vertex from
Eq.~(\ref{FF_OPE_3_zeromomentum}) can be rewritten as   
\be
\lim_{\lambda \to \infty} {\cal F}_{{\pi^0}^*\gamma^*\gamma}((\lambda
q_1)^2, (\lambda q_1)^2,0) = {F_0 \over 3} \ \chi 
+ \order{{1\over \lambda}} .
\label{FF_OPE_3_zeromomentum_chi}  
\ee

Unfortunately there is no agreement in the literature what the actual value of
$\chi$ should be. In comparing different results one has to keep in mind that
$\chi$ actually depends on the renormalization scale $\mu$. In
Ref.~\cite{Ioffe_Smilga} the estimate $\chi(\mu = 0.5~\mbox{GeV}) = -
(8.16^{+2.95}_{-1.91})~\mbox{GeV}^{-2}$ was given in a QCD sum rule evaluation
of nucleon magnetic moments. This value was confirmed by the recent
reanalysis~\cite{Narison:2008jp} which yields $\chi = - (8.5 \pm
1.0)~\mbox{GeV}^{-2}$, although no scale $\mu$ has been specified. A similar
value $\chi = - N_C / (4 \pi^2 F_\pi^2) = - 8.9~\mbox{GeV}^{-2}$ was obtained
in Ref.~\cite{Vainshtein03}. From the explicit expression of $\chi$ it is not
immediately clear what should be the relevant scale $\mu$. Since pion
dominance was used in the matching with the OPE below some higher states, it
was argued in Ref.~\cite{Vainshtein03} that the normalization point is
probably rather low, $\mu \sim 0.5~\mbox{GeV}$.  Calculations within the
instanton liquid model yield $\chi^{\rm ILM}(\mu \sim 0.5-0.6~\mbox{GeV}) =
-4.32~\mbox{GeV}^{-2}$~\cite{Chi_ILM_1}, where the scale is set by the inverse
average instanton size $\rho^{-1}$. The value of $\chi
\langle{\overline\psi}\psi\rangle_0 = 42~\mbox{MeV}$ at the same scale
obtained in Ref.~\cite{Chi_ILM_1} agrees roughly with the result
$35-40~\mbox{MeV}$ from Ref.~\cite{Chi_ILM_2} derived in the same model.

The leading short-distance behavior of the two-point function $\Pi_{\rm VT}$
in Eq.~(\ref{Pi_VT_invariant_function}) is given by~\cite{KN_EPJC01} (see also
Ref.~\cite{Craigie:1981jx})
\be \label{VT_OPE} 
\lim_{\lambda \to \infty}\Pi_{\rm VT}((\lambda p)^2)\,=\,
-\,\frac{1}{\lambda^2}
\,\frac{\langle{\overline\psi}\psi\rangle_0}{p^2}
\,+\,{\cal O}\left(\frac{1}{\lambda^4}\right)\,. 
\ee
Assuming that $\Pi_{\rm VT}(p^2)$ is well described by the multiplet of the
lowest-lying vector mesons (LMD) and satisfies the OPE constraint from
Eq.~(\ref{VT_OPE}) leads to the
Ansatz~\cite{Balitsky_Yung,Belyaev_Kogan,KN_EPJC01}
\be \label{Pi_VT_LMD}
\Pi_{\rm VT}^{\rm LMD}(p^2) \,=\, -\,\langle{\overline\psi}\psi\rangle_0\,
\frac{1}{p^2-M_V^2} \, .  
\ee
Using Eq.~(\ref{Pi_VT0_Chi}) then leads to the estimate $\chi^{\rm LMD} = - 2
/ M_V^2 = -3.3~\mbox{GeV}^{-2}$~\cite{Balitsky_Yung}. Again, it is not obvious
at which scale this relation holds. In analogy to estimates of low-energy
constants in chiral Lagrangians~\cite{Ecker_etal}, it might be at $\mu = M_V$,
although in principle the renormalization scale of $\chi$ is not related to
the one of low-energy constants; see the discussion in Ref.~\cite{Cata_Mateu}.
This LMD estimate was soon afterwards improved by taking into account higher
resonance states ($\rho^\prime, \rho^{\prime\prime}$) in the framework of QCD
sum rules, with the results $\chi(0.5~\mbox{GeV}) = - (5.7 \pm
0.6)~\mbox{GeV}^{-2}$~\cite{Belyaev_Kogan} and $\chi(1~\mbox{GeV}) = - (4.4
\pm 0.4)~\mbox{GeV}^{-2}$~\cite{Balitsky_etal}. A more recent
analysis~\cite{Ball_etal} yields, however, a smaller absolute value
$\chi(1~\mbox{GeV}) = - (3.15 \pm 0.30)~\mbox{GeV}^{-2}$, close to the
original LMD estimate. Further arguments for the latter value are also given
in Ref.~\cite{Mateu_Portoles} and references therein, by studying the coupling
of the tensor current to the $\rho$ meson.  For a quantitative comparison of
all these estimates for $\chi$ we would have to run them to a common scale,
for instance, 1 GeV, which can obviously not be done within perturbation
theory starting from such low scales as $\mu = 0.5~\mbox{GeV}$.\footnote{A
further complication arises in comparisons with papers from the early 1980's
because not only $\mu = 0.5~\mbox{GeV}$ was frequently used, but also 1-loop
running with a low $\Lambda_{\rm QCD}^{n_f=3} = 100-150~\mbox{MeV}$, whereas
more recent estimates yield $\Lambda_{\overline{\rm MS}}^{n_f=3} =
346~\mbox{MeV}$ (at 4-loop)~\cite{Bethke_2000}.} Finally, even if the
renormalization-group running could be performed nonperturbatively, it is not
clear what would be the relevant scale $\mu$ in the context of hadronic
light-by-light scattering.

\begin{figure}[h]
\centering
\centerline{\epsfig{figure=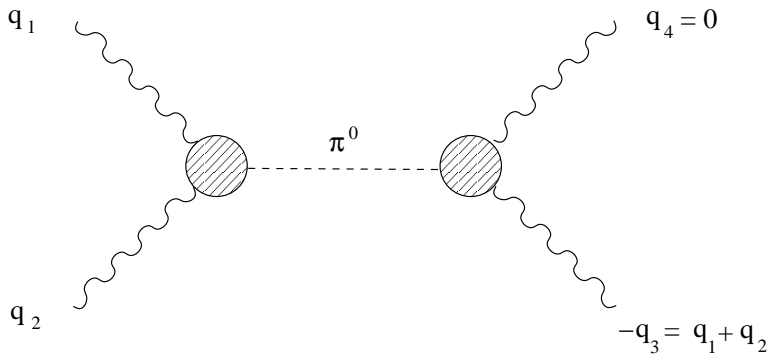,height=4.6cm,width=10cm}}
\caption{Pion exchange in the $s$ channel in hadronic light-by-light
  scattering. The photon with zero momentum $q_4 = 0$ represents the external
  soft photon for the corresponding contribution to the muon $g-2$.} 
\label{fig:pionexchange-s-channel}
\end{figure}

A short-distance constraint on the pion-exchange contribution to the hadronic
light-by-light scattering correction in the muon $g-2$ itself was derived in
Ref.~\cite{MV03}. The relevant kinematical configuration for the $s$-channel
exchange of the pion is shown in Fig.~\ref{fig:pionexchange-s-channel}. 
In general one has $q_1 + q_2 + q_3 + q_4 = 0$, but for the muon $g-2$ the
soft photon limit $q_4 \to 0$ will be relevant.  The authors of
Ref.~\cite{MV03} then consider the limit $q_1^2 \sim q_2^2 \gg q_3^2$, where
$q_3 = - (q_1 + q_2)$. Since in this limit the leading term in the OPE of the
two vector currents associated with the momenta $q_1$ and $q_2$ yields the
axial-vector current, they can relate the matrix element $\langle VVV | \gamma
\rangle$ which enters for the muon $g-2$ to the famous anomalous triangle
diagram $\langle A V | \gamma\rangle$~\cite{ABJanomaly}, which is highly
constrained; see Refs.~\cite{nonrenormalization,Vainshtein03}.  From this they
deduce that no momentum-dependent form factor should be used at the external
vertex, but only a constant factor. They thus obtain the following
intermediate expression for the light-by-light scattering
amplitude\footnote{We have rescaled the form factor in Eq.~(18) in
Ref.~\cite{MV03} to agree with our normalization in
Eq.~(\ref{FF_normalization_WZW}) and used Minkowski space notation.}: 
\be \label{A_pi0_MV}
{\cal A}_{\pi^0} = {3 \over 2 F_\pi} \, 
{{\cal F}_{\pi^0\gamma^*\gamma^*}(q_1^2, q_2^2) \over q_3^2 - m_\pi^2} \,
(f_{2; \mu\nu} \tilde f_1^{\nu\mu}) (\tilde f_{\rho\sigma} f_3^{\sigma\rho})  
\ + \ \mbox{permutations}, 
\ee
where $f_i^{\mu\nu} = q_i^\mu \epsilon_i^\nu - q_i^\nu \epsilon_i^\mu$ and
$\tilde f_{i; \mu\nu} = {1 \over 2} \epsilon_{\mu\nu\rho\sigma}
f_i^{\rho\sigma}$ for $i=1,2,3$ denote the field strength tensors of the
internal photons with polarization vectors $\epsilon_i$. The field strength
tensor of the external soft photon is defined similarly by $f^{\mu\nu} =
q_4^\mu \epsilon_4^\nu - q_4^\nu \epsilon_4^\mu$. Except in $\tilde
f_{\rho\sigma}$ the limit $q_4 \to 0$ is understood in Eq.~(\ref{A_pi0_MV}),
in particular, in $f_3^{\sigma\rho}$ and in the pion propagator.

Note the absence of a second form factor ${\cal
  F}_{\pi^0\gamma^*\gamma}(q_3^2, 0)$ in Eq.~(\ref{A_pi0_MV}) at the external
  vertex.  The authors of Ref.~\cite{MV03} rightly point out that such a
  momentum-dependent form factor at the external vertex would violate momentum
  conservation and criticize the procedure adopted in earlier
  works~\cite{BPP95,HKS95,HK98,KN01}.  However, it is obvious from their
  expressions [Eq.~(18) in Ref.~\cite{MV03}], reproduced here in
  Eq.~(\ref{A_pi0_MV}), that they only consider the on-shell pion form factor
  ${\cal F}_{{\pi^0}\gamma^*\gamma^*}(q_1^2, q_2^2) \equiv {\cal
  F}_{{\pi^0}\gamma^*\gamma^*}(m_\pi^2, q_1^2, q_2^2)$ at the internal vertex
  and not the off-shell pion form factor ${\cal
  F}_{{\pi^0}^*\gamma^*\gamma^*}(q_3^2, q_1^2, q_2^2)$.  Note that the
  expression in Eq.~(\ref{A_pi0_MV}) has to be compared with the term
  involving $T_2$ in Eq.~(\ref{a_pion_2}).  Therefore, contrary to the claim
  in Ref.~\cite{MV03}, they only consider the {\it pion-pole} contribution to
  hadronic light-by-light scattering and not the {\it pion-exchange}
  contribution which involves fully off-shell form factors at the internal and
  the external vertex. Actually, also a second argument in Ref.~\cite{MV03}
  [after Eq.~(20) there] in favor of a constant form factor at the external
  vertex is clearly based on the use of on-shell form factors. The use of a
  nonconstant on-shell form factor ${\cal
  F}_{{\pi^0}\gamma^*\gamma}(q_3^2,0)$ at the external vertex would lead,
  together with the pion propagator, to an overall $1/q_3^4$ behavior, since
  ${\cal F}_{{\pi^0}\gamma^*\gamma}(q_3^2,0) \sim 1/q_3^2$, for large $q_3^2$,
  according to Brodsky-Lepage; see Eq.~(\ref{Brodsky_Lepage}). This would
  contradict the $1/q_3^2$ behavior observed in Eq.~(\ref{A_pi0_MV}) (apart
  from $f_3^{\sigma\rho}$).

\section{New evaluation of the pseudoscalar-exchange con\-tri\-bu\-tion in
  large-$N_C$ QCD}
\label{sec:new_evaluation_pseudoscalars}

In the spirit of the minimal hadronic Ansatz for Green's functions in
large-$N_C$ QCD, on-shell ${\cal F}_{{\pi^0}\gamma^*\gamma^*}(m_\pi^2, q_1^2,
q_2^2)$ and off-shell form factors ${\cal
F}_{{\pi^0}^*\gamma^*\gamma^*}((q_1+q_2)^2, q_1^2, q_2^2)$ have been
constructed in Ref.~\cite{KN_EPJC01}. They contain either the lowest-lying
multiplet of vector resonances (LMD) or two multiplets, the $\rho$ and the
$\rho'$ (LMD+V). Both Ans\"atze fulfill {\it all} the OPE constraints from
Eqs.~(\ref{FF_OPE_1}), (\ref{FF_OPE_2}) and (\ref{FF_OPE_3}); however, the LMD
Ansatz does {\it not} reproduce the Brodsky-Lepage behavior from
Eq.~(\ref{Brodsky_Lepage}). Instead it behaves like ${\cal
F}_{{\pi^0}\gamma^*\gamma}^{\rm LMD}(m_\pi^2, -Q^2, 0) \sim
\mbox{const}$. The $1/Q^2$ falloff can be achieved with the LMD+V Ansatz with
a certain choice of the free parameters; see below.  Note that it might not
always be possible to satisfy all short-distance constraints, in particular
from the high-energy behavior of form factors, if only a finite number of
resonances is included; see Ref.~\cite{BGLP03}.  The on-shell form factors
were later used in Ref.~\cite{KN01} to evaluate the pion-pole contribution;
see also Ref.~\cite{Bijnens_Persson01}. 

In the following, we reevaluate the pion-exchange contribution using {\it
  off-shell} LMD+V form factors~\cite{KN_EPJC01} at both vertices
\bea
{\cal F}_{{\pi^0}^*\gamma^*\gamma^*}^{\rm LMD+V}(q_3^2, q_1^2, q_2^2)&=&
 \frac{F_\pi}{3}\, {q_1^2\,q_2^2\,(q_1^2 + q_2^2 + q_3^2) + P_H^V(q_1^2,
   q_2^2, q_3^2) \over (q_1^2-M_{V_1}^2) \, (q_1^2-M_{V_2}^2) \,
   (q_2^2-M_{V_1}^2) \, (q_2^2-M_{V_2}^2)} , 
\label{KNpipioff} \\  
P_H^V(q_1^2,q_2^2,q_3^2)&=& h_1\,(q_1^2+q_2^2)^2
+ h_2\,q_1^2\,q_2^2 + h_3\,(q_1^2+q_2^2)\,q_3^2 + h_4\,q_3^4 \nonumber \\  
&& +h_5\,(q_1^2+q_2^2) + h_6\,q_3^2 + h_7, 
\eea
with $q_3^2 = (q_1 + q_2)^2$. In the spirit of resonance chiral
theory~\cite{Ecker_etal} a Lagrangian with two multiplets of vector resonances
was proposed recently in Ref.~\cite{Mateu_Portoles} and references therein,
which reproduces the above LMD+V Ansatz and which fulfills all the QCD
short-distance constraints for the $\VVP$ Green's function.

The constants $h_i$ in the Ansatz for ${\cal F}^{\rm LMD+V}_{{\pi^0}^*
\gamma^* \gamma^*}$ in Eq.~(\ref{KNpipioff}) are determined as follows. The
normalization with the pion decay amplitude $\pi^0 \to \gamma\gamma$ in
Eq.~(\ref{FF_normalization_WZW}) yields $h_7 = - N_C M_{V_1}^4 M_{V_2}^4 / (4
\pi^2 F_\pi^2) - h_6 m_\pi^2 - h_4 m_\pi^4 = -14.83~\mbox{GeV}^6 - h_6 m_\pi^2
- h_4 m_\pi^4$, where we used $M_{V_1} = M_\rho = 775.49~\mbox{MeV}$ and
$M_{V_2} = M_{\rho^\prime} = 1.465~\mbox{GeV}$~\cite{PDG2008}. Note that in
Refs.~\cite{KN_EPJC01,KN01} the small corrections proportional to the pion
mass were dropped, assuming that the $|h_i|$ are of order $1-10$ in appropriate
units of GeV.  The Brodsky-Lepage behavior from Eq.~(\ref{Brodsky_Lepage}) can
be reproduced by choosing $h_1 = 0~\mbox{GeV}^2$. Furthermore, in
Ref.~\cite{KN_EPJC01} a fit to the CLEO data for the on-shell form factor
${\cal F}_{{\pi^0}\gamma^*\gamma}^{\rm LMD+V}(m_\pi^2, -Q^2, 0)$ was
performed, with the result $h_5 = (6.93 \pm 0.26)~\mbox{GeV}^4 - h_3
m_\pi^2$. Again, the correction proportional to the pion mass was omitted in
Refs.~\cite{KN_EPJC01,KN01}. As pointed out in Ref.~\cite{MV03}, the constant
$h_2$ can be obtained from the higher-twist corrections in the OPE. Comparing
with Eq.~(\ref{NSVVZ}) yields the result $h_2=-4\,(M_{V_1}^2+M_{V_2}^2) +
(16/9)\,\delta^2 \simeq -10.63~\gev^2$.

Within the LMD+V framework, the vector-tensor two-point function
reads~\cite{KN_EPJC01}  
\bea
\Pi_{\rm VT}^{\rm LMD+V}(p^2) & = &  -\,\langle{\overline\psi}\psi\rangle_0\, 
{ p^2 + c_{\rm VT} \over (p^2-M_{V_1}^2) (p^2-M_{V_2}^2) } \, ,
\label{VT_LMD+V} \\ 
c_{\rm VT} & = & {M_{V_1}^2 M_{V_2}^2 \chi \over 2} , 
\label{c_VT}
\eea
where we fixed the constant $c_{\rm VT}$ using Eq.~(\ref{Pi_VT0_Chi}). As
shown in Ref.~\cite{KN_EPJC01} the OPE from Eq.~(\ref{FF_OPE_3}) for ${\cal
  F}_{{\pi^0}^*\gamma^*\gamma^*}^{\rm LMD+V}$ leads to the relation
\be \label{constraint_h1_h3_h4} 
h_1 + h_3 + h_4 = 2 c_{\rm VT} . 
\ee
As noted above, the value of the magnetic susceptibility $\chi(\mu)$ and the
relevant scale $\mu$ are not precisely known. However, the LMD estimate
$\chi^{\rm LMD} = - 2 / M_V^2 = -3.3~\mbox{GeV}^{-2}$ is close to $\chi(\mu =
1~\mbox{GeV}) = -(3.15 \pm 0.30)~\mbox{GeV}^{-2}$ obtained in
Ref.~\cite{Ball_etal} using QCD sum rules with several vector resonances
$\rho, \rho^\prime$, and $\rho^{\prime\prime}$. Assuming that the LMD/LMD+V
framework is self-consistent, we will therefore take $\chi = (-3.3 \pm
1.1)~\mbox{GeV}^{-2}$ in our numerical evaluation, with a typical large-$N_C$
uncertainty of about 30\%.  This translates into the constraint $h_3 + h_4 =
(-4.3 \pm 1.4)~\mbox{GeV}^{2}$, corresponding to $c_{\rm VT} = (-2.13 \pm
0.71)~\mbox{GeV}^2$. We will vary $h_3$ in the range $\pm 10~\mbox{GeV}^2$ and
determine $h_4$ from Eq.~(\ref{constraint_h1_h3_h4}) and vice versa.

Note that using the off-shell LMD+V form factor at the external vertex leads
to a short-distance behavior in the full light-by-light scattering
contribution which at least qualitatively agrees with the OPE constraint
derived in Ref.~\cite{MV03}.  As stressed earlier, Ref.~\cite{MV03} only
considers the pion-pole contribution with on-shell form factors; therefore a
direct quantitative comparison with our approach is not
possible. Nevertheless, taking first $q_1^2 \sim q_2^2 \gg q_3^2$ and then
$q_3^2$ large, one obtains, together with the pion propagator in
Eq.~(\ref{a_pion_2}) [in the term with $T_2$], an overall $1/q_3^2$ behavior
for the pion-exchange contribution, since at the external vertex we
have~\cite{Jegerlehner_private_communication}
\be \label{external_vertex}
\frac{3}{F_\pi}\,{\cal F}_{{\pi^0}^*\gamma^*\gamma}^{\rm LMD+V}(q_3^2, q_3^2,
0)\stackrel{q_3^2 \to \infty}{\to}\frac{h_1+h_3+h_4}{M_{V_1}^2
  M_{V_2}^2}=\frac{2c_{\rm VT}}{M_{V_1}^2 M_{V_2}^2}=\chi. 
\ee
In the derivation we used Eqs.~(\ref{c_VT}) and (\ref{constraint_h1_h3_h4}),
see also Eq.~(\ref{FF_OPE_3_zeromomentum_chi}).  This $1/q_3^2$ behavior is as
expected from Eq.~(\ref{A_pi0_MV}), reproduced earlier from Ref.~\cite{MV03}.
On the other hand, if we would use a constant form factor proportional to the
WZW term at the external vertex as proposed in Ref.~\cite{MV03}, we would
get~\cite{Jegerlehner_private_communication}
\be
\frac{3}{F_\pi}\,{\cal F}_{{\pi^0}\gamma\gamma}^{\rm LMD+V}(0, 0, 0)=
\frac{h_7}{M_{V_1}^4M_{V_2}^4}=-\frac{N_C}{4\pi^2F_\pi^2}\simeq
-8.9~\gev^{-2},  
\ee
where for simplicity we considered the chiral limit.  That means that with the
value of $\chi = - N_C / (4 \pi^2 F_\pi^2)$ from Ref.~\cite{Vainshtein03} we
would in Eq.~(\ref{external_vertex}) precisely satisfy the short-distance
constraint from Ref.~\cite{MV03}. 

The coefficient $h_6$ in the LMD+V Ansatz is undetermined as well. We can
obtain some indirect information on its size and sign in the following
way. Low-energy constants in chiral Lagrangians can be estimated by starting
with some resonance Lagrangian and then integrating out the heavy resonance
states, usually at tree level. In particular for low-energy constants which
are given by the exchanges of vector and axial-vector mesons, this procedure
works in general quite well~\cite{Lres_LEC, Ecker_etal}. Although for instance
for vector mesons one can write down many different Lagrangians, it was shown
in Ref.~\cite{Ecker_etal} that imposing QCD short-distance constraints on the
resonance Lagrangian itself leads to unique estimates for the low-energy
constants at order $p^4$ in the chiral Lagrangian. At order $p^6$ this is not
true anymore~\cite{MoussallamStern,KN_EPJC01}; nevertheless, it seems
reasonable to reduce the model dependence by imposing again short-distance
constraints on such resonance Lagrangians.

Usually, only the exchange of the lightest resonance state in each channel is
considered in this approach. One expects, however, some corrections to these
estimates, with a typical large-$N_C$ error of about 30\%, if also the
exchanges of heavier resonance states are taken into account.  In
Ref.~\cite{KN_EPJC01} this was shown to be true for one of two linear
combinations of low-energy constants from the chiral Lagrangian of odd
intrinsic parity at order $p^6$ which enter in the low-energy expansion of the
Green's function $\VVP$. With the LMD and LMD+V Ans\"atze for this Green's
function, the relevant combination of low-energy constants is given by
\bea
A_{V,p^2}^{\rm LMD} & = & {F_\pi^2 \over 8 M_V^4} - {N_C \over 32 \pi^2 M_V^2}
= -1.11 \ {10^{-4} \over F_\pi^2}, \label{A_Vp2_LMD} \\ 
A_{V,p^2}^{\rm LMD+V} & = & {F_\pi^2 \over 8 M_{V_1}^4} {h_5 \over M_{V_2}^4}
- {N_C \over 32 \pi^2 M_{V_1}^2} \left(1 + {M_{V_1}^2 \over M_{V_2}^2}
\right) = -1.36 \ {10^{-4} \over F_\pi^2}. \label{A_Vp2_LMD_V} 
\eea
The constant $h_5$ which enters $A_{V,p^2}^{\rm LMD+V}$ is directly related to
the Brodsky-Lepage behavior of the form factor. Even though this falloff
behavior cannot be reproduced with the LMD Ansatz, the change in the
low-energy constant when going from LMD to LMD+V is only about 20\%, well
within the expected large-$N_C$ uncertainty.

On the other hand, the coefficient $h_6$ determines a second linear
combination of low-energy constants at order $p^6$: 
\be \label{A_V_pq}
A_{V,(p+q)^2}^{\rm LMD} = - {F_\pi^2 \over 8 M_V^4} = -0.26 \ {{10^{-4} \over
    F_\pi^2}}, \qquad \qquad 
A_{V,(p+q)^2}^{\rm LMD+V} = - {F_\pi^2 \over 8 M_{V_1}^4 M_{V_2}^4} h_6. 
\ee
Note that using the resonance Lagrangian of Ref.~\cite{Prades94}, one would
obtain $A_{V,(p+q)^2}^{\rm res} = 0$ instead. However, this resonance
Lagrangian in general fails to reproduce the short-distance constraints from
QCD, in contrast to the LMD and LMD+V Ans\"atze; see Ref.~\cite{KN_EPJC01}. In
particular, the prediction for $A_{V,(p+q)^2}$ in the LMD model follows
directly from the implementation of these short-distance constraints.  Note
that there is no problem with the short-distance behavior for the LMD form
factor in the relevant channel where at low energies $A_{V,(p+q)^2}$
enters. If we would assume a 30\% error on the LMD estimate in
Eq.~(\ref{A_V_pq}), we would obtain the quite narrow range $h_6 = M_{V_2}^4 (1
\pm 0.3) = (4.6 \pm 1.4)~\mbox{GeV}^4$.  However, this procedure might
underestimate the potential variation of $h_6$, since the low-energy constant
$A_{V,(p+q)^2}^{\rm LMD}$ happens to be small compared to $A_{V,p^2}^{\rm
LMD}$; see Eq.~(\ref{A_Vp2_LMD}). The magnitude of the shift of $A_{V,p^2}$
when going from LMD to LMD+V is $-0.25 \ (10^{-4} / F_\pi^2)$. That is, the
shift is of the same size as $A_{V,(p+q)^2}^{\rm LMD}$ itself. Assuming again
that the LMD/LMD+V framework is self-consistent, but, to be conservative,
allowing for a 100\% uncertainty of $A_{V,(p+q)^2}^{\rm LMD}$, we get the
range $h_6 = (5 \pm 5)~\mbox{GeV}^4$.

Of course, the uncertainties of the values of the undetermined parameters
$h_3, h_4$ and $h_6$ and of the magnetic susceptibility $\chi(\mu)$ is a
drawback when using the off-shell LMD+V form factor and will limit the
precision of the final estimate.

The integral to be performed in Eq.~(\ref{a_pion_2}) is eight-dimensional,
thereof 3 integrations can be done trivially. In general, one then has to deal
with a five-dimensional integration over 3 angles and 2 moduli. We have
performed these integrations numerically after a rotation to Euclidean momenta
using the program VEGAS~\cite{VEGAS}.  As a check we have reproduced the
values of $a_{\mu}^{\mathrm{LbyL};\pi^0}$ for various form factors which have
been used earlier in the literature. For instance, using a simple VMD form
factor, we obtain $a_{\mu; \mathrm{VMD}}^{\mathrm{LbyL};\pi^0} = 57 \times
10^{-11}$ for $m_\mu = 105.658369~\mev$ and $m_{\pi^0} = 134.9766~\mev$ and
with the value $M_\rho = 775.49~\mev$.

The results for $a_\mu^{\mathrm{LbyL};\pi^0}$ for some selected values of $h_3,
h_4$ and $h_6$, varied in the ranges discussed above, for $\chi =
-3.3~\mbox{GeV}^{-2}$, $h_1 = 0~\mbox{GeV}^2$, $h_2 = -10.63~\mbox{GeV}^2$ and
$h_5 = 6.93~\mbox{GeV}^4 - h_3 m_\pi^2$ are collected in
Table~\ref{tab:pi0res}. 
\begin{table}[h] 
\caption{Results for $a_\mu^{\mathrm{LbyL};\pi^0}\times 10^{11}$ obtained with
the off-shell LMD+V form factor for $\chi = -3.3~\mbox{GeV}^{-2}$ and the
given values for $h_3, h_4$ and $h_6$. When varying $h_3$ (upper half of the
table), the parameter $h_4$ is fixed by the constraint in
Eq.~(\ref{constraint_h1_h3_h4}). In the lower half the procedure is
reversed. The values of the other parameters are given in the text.} 
\begin{center} 
\renewcommand{\arraystretch}{1.25}
\begin{tabular}{|l|c|c|c|}
\hline 
 & ~$h_6 = 0~\mbox{GeV}^4$~ & ~$h_6 = 5~\mbox{GeV}^4$~ & ~$h_6 =
10~\mbox{GeV}^4$~ \\  
\hline 
~$h_3 = -10~\mbox{GeV}^2$~ & 68.4 & 74.1 & 80.2 \\ 
~$h_3 = 0~\mbox{GeV}^2$    & 66.4 & 71.9 & 77.8 \\
~$h_3 = 10~\mbox{GeV}^2$   & 64.4 & 69.7 & 75.4 \\ 
\hline 
~$h_4 = -10~\mbox{GeV}^2$  & 65.3 & 70.7 & 76.4 \\
~$h_4 = 0~\mbox{GeV}^2$    & 67.3 & 72.8 & 78.8 \\
~$h_4 = 10~\mbox{GeV}^2$   & 69.2 & 75.0 & 81.2 \\
\hline 
\end{tabular}
\label{tab:pi0res}
\end{center} 
\end{table}

Varying $\chi$ by $\pm 1.1~\mbox{GeV}^{-2}$ changes the result for
$a_\mu^{\mathrm{LbyL};\pi^0}$ by $\pm 2.1 \times 10^{-11}$ at most.  One
observes from the table that the uncertainty in $h_6$ affects the result by up
to $\pm 6.4 \times 10^{-11}$. If we would use instead $h_6 = (0 \pm
10)~\mbox{GeV}^4$, the result would vary by about $\pm 12 \times 10^{-11}$
around the central value. The variation of $a_\mu^{\mathrm{LbyL};\pi^0}$ with
$h_3$ [with $h_4$ determined from the constraint in
Eq.~(\ref{constraint_h1_h3_h4}) or vice versa] is much smaller, at most $\pm
2.5 \times 10^{-11}$. The variation of $h_5$ by $\pm 0.26~\mbox{GeV}^4$ only
leads to changes of $\pm 0.6 \times 10^{-11}$ in the final result.  Within the
scanned region, we obtain a minimal value of $a_\mu^{\mathrm{LbyL};\pi^0} =
63.2 \times 10^{-11}$ for $\chi = -2.2~\mbox{GeV}^{-2}, h_3 =
10~\mbox{GeV}^2$, and $h_6 = 0~\mbox{GeV}^4$ and a maximum of
$a_\mu^{\mathrm{LbyL};\pi^0} = 83.3 \times 10^{-11}$ for $\chi =
-4.4~\mbox{GeV}^{-2}, h_4 = 10~\mbox{GeV}^2$, and $h_6 = 10~\mbox{GeV}^4$.  In
the absence of more information on the precise values of the constants $h_3,
h_4$ and $h_6$, we take the average of the results obtained with $h_6 =
5~\mbox{GeV}^4$ for $h_3 = 0~\mbox{GeV}^2$, i.e.\ $71.9 \times 10^{-11}$, and
for $h_4 =0~\mbox{GeV}^2$, i.e.\ $72.8 \times 10^{-11}$, as our central value,
$72.3 \times 10^{-11}$. To estimate the error, we add all the uncertainties
from the variations of $\chi$, $h_3$ (or $h_4$), $h_5$ and $h_6$ linearly to
cover the full range of values obtained with our scan of parameters. Note that
the uncertainties of $\chi$ and the coefficients $h_3,h_4$ and $h_6$ do not
follow a Gaussian distribution. In this way we obtain our final estimate
\be \label{amupi0LMD+V}
a_\mu^{\mathrm{LbyL};\pi^0} = (72 \pm 12) \times 10^{-11}.
\ee
We think the 16\% error should fairly well describe the inherent model
uncertainty using the {\it off-shell} LMD+V form factor. In order to
facilitate future updates of our result in case some of the parameters $h_i$
in the LMD+V Ansatz in Eq.~(\ref{KNpipioff}) or the value (and the relevant
scale) of the magnetic susceptibility $\chi(\mu)$ will be known more
precisely, we present in the appendix a parametrization of 
$a_\mu^{\mathrm{LbyL};\pi^0}$ for arbitrary coefficients $h_i$.

As far as the contribution to $a_\mu$ from the exchanges of the other light
pseudoscalars $\eta$ and $\eta^\prime$ is concerned, it is not so
straightforward to apply the above analysis within the LMD+V framework to
these resonances. In particular, the short-distance analysis in
Ref.~\cite{KN_EPJC01} was performed in the chiral limit and assumed octet
symmetry. For the $\eta$ the effect of nonzero quark masses has definitely to
be taken into account. Furthermore, the $\eta^\prime$ has a large admixture
from the singlet state and the gluonic contribution to the axial anomaly will
play an important role. We therefore resort to a simplified approach which was
also adopted in other works~\cite{BPP95,HKS95,HK98,KN01,MV03}
and take a simple VMD form factor
\be
{\cal F}_{{\rm PS}^*\gamma^*\gamma^*}^{\rm VMD}(q_3^2,q_1^2,q_2^2) = -
\frac{N_C }{ 12 \pi^2 F_{\rm PS}} \frac{M_V^2 }{(q_1^2 - M_V^2)} \frac{M_V^2
}{ (q_2^2 - M_V^2)}, \qquad \mbox{PS} = \eta, \eta^\prime, 
\label{FF_VMD}
\ee
normalized to the experimental decay width $\Gamma(\mbox{PS} \to \gamma
\gamma)$. We can fix the normalization by adjusting the (effective)
pseudoscalar decay constant $F_{\rm PS}$ in Eq.~(\ref{FF_VMD}). Using the
latest values $\Gamma(\eta \to \gamma \gamma) = (0.510 \pm 0.026)~\mbox{keV}$
and $\Gamma(\eta^\prime \to \gamma \gamma) = (4.30 \pm 0.15)~\mbox{keV}$ from
Ref.~\cite{PDG2008}, one obtains $F_{\eta, {\rm eff}} = 93.0~\mbox{MeV}$ with
$m_\eta = 547.853~\mbox{MeV}$ and $F_{\eta^\prime, {\rm eff}} =
74.0~\mbox{MeV}$ with $m_{\eta^\prime} = 957.66~\mbox{MeV}$.  However, we do
not follow the approach of Ref.~\cite{MV03} and will also take a VMD form
factor at the external vertex.

Note that the on- and off-shell VMD form factors are identical, since the form
factor does not depend on the momentum $q_3^2$ which flows through the pion
leg.  The problem with the VMD form factor is that the damping is too strong
as it behaves like ${\cal F}_{{\pi^0}\gamma^*\gamma^*}(m_\pi^2, -Q^2, -Q^2)
\sim 1/Q^4$, instead of $\sim 1/Q^2$ deduced from the OPE, see
Eq.~(\ref{FF_OPE_2}).  This effect might lead to an underestimating of the
contribution. However, the relevant integrals for $a_\mu^{\mathrm{LbyL; PS}}$
do not seem to be very sensitive to the correct asymptotic behavior for large
momenta. This can be seen from the weight functions which multiply the form
factors in the integral and which are displayed in Ref.~\cite{KN01}. It seems
more important to have a good description at small and intermediate energies
below 1~GeV, e.g.\ by reproducing the slope of the form factor ${\cal
F}_{\mathrm{PS}\gamma^*\gamma}(-Q^2,0)$ at the origin. The CLEO
Collaboration~\cite{CLEO98} has made a fit of the on-shell form factors ${\cal
F}_{\eta\gamma^*\gamma}(-Q^2,0)$ and ${\cal F}_{\eta^\prime
\gamma^*\gamma}(-Q^2,0)$, normalized to the corresponding experimental width
$\Gamma(\mbox{PS} \to \gamma \gamma)$, using a VMD Ansatz with an adjustable
parameter $\Lambda_{\mathrm{PS}}$ in place of the vector-meson mass $M_V$ in
Eq.~(\ref{FF_VMD}). Taking their values $\Lambda_\eta = (774 \pm
29)~\mbox{MeV}$ and $\Lambda_{\eta^\prime} = (859 \pm 28)~\mbox{MeV}$, we then
obtain the results $a_{\mu}^{\mathrm{LbyL};\eta} = 14.5 \times 10^{-11}$ and
$a_{\mu}^{\mathrm{LbyL};\eta^\prime} = 12.5 \times 10^{-11}$, which update the
values given in Ref.~\cite{KN01}.\footnote{If we use a constant (WZW) form
factor at the external vertex, as proposed in Ref.~\cite{MV03}, we would
obtain $a_{\mu}^{\mathrm{LbyL};\eta-\mathrm{pole}} = 21.5 \times 10^{-11}$ and
$a_{\mu}^{\mathrm{LbyL};\eta^\prime-\mathrm{pole}} = 20.1 \times
10^{-11}$. Note that these values are somewhat larger than
$a_{\mu}^{\mathrm{LbyL};\eta-\mathrm{pole}} = 18 \times 10^{-11}$ and
$a_{\mu}^{\mathrm{LbyL};\eta^\prime-\mathrm{pole}} = 18 \times 10^{-11}$ given
in Ref.~\cite{MV03}.} Only a more detailed analysis, along the line of the
LMD+V framework, will show whether these values are realistic. Thus, adding up
the contributions from all the light pseudoscalar exchanges ($\pi^0, \eta,
\eta^\prime$), we obtain the estimate
\be \label{amuLbLPS}
a_{\mu}^{\mathrm{LbyL;PS}} = (99 \pm 16) \times 10^{-11}, 
\ee
where we have assumed a 16\% error, as inferred above for pion-exchange
contribution using the off-shell LMD+V form factor.\footnote{Applying the same
procedure to the electron, we obtain $a_e^{\mathrm{LbyL};\pi^0} = (2.98 \pm
0.34) \times 10^{-14}$ with off-shell LMD+V form factors at both
vertices. This number supersedes the value given in
Ref.~\cite{KN01}. Note that the naive rescaling
$a_e^{\mathrm{LbyL};\pi^0}(\mathrm{rescaled}) = (m_e / m_\mu)^2
a_\mu^{\mathrm{LbyL};\pi^0} = 1.7 \times 10^{-14}$ yields a value which is
almost a factor of 2 too small. Our estimates for the other pseudoscalars
contributions using VMD form factors at both vertices are
$a_e^{\mathrm{LbyL};\eta} = 0.49 \times 10^{-14}$ and
$a_e^{\mathrm{LbyL};\eta^\prime} = 0.39 \times 10^{-14}$. Therefore we get
$a_e^{\mathrm{LbyL;PS}} = (3.9 \pm 0.5) \times 10^{-14}$, where the relative
error of about 12\% is again taken over from the pion-exchange contribution.}

\section{Discussion and conclusions}
\label{sec:conclusions} 
 
Following the observation in Refs.~\cite{FJ_Essentials, FJ_Book} we have
reevaluated the pion-exchange contribution to hadronic light-by-light
scattering in the muon $g-2$ using fully off-shell form factors ${\cal
F}_{{\pi^0}^*\gamma^*\gamma^*}$ at both vertices. We used a model based on the
large-$N_C$ QCD framework with two multiplets of vector mesons
(LMD+V)~\cite{KN_EPJC01} which fulfills all QCD short-distance constraints on
the form factor and reproduces the experimentally confirmed Brodsky-Lepage
behavior. We also derived a new short-distance constraint on the form factor
at the external vertex, relating it to the quark condensate magnetic
susceptibility $\chi$. The obtained value $a_\mu^{\mathrm{LbyL};\pi^0} = (72
\pm 12) \times 10^{-11} $ replaces the result obtained in Ref.~\cite{KN01}
with on-shell LMD+V form factors at both vertices. Adding the contribution
from the exchanges of the $\eta$ and $\eta^\prime$ evaluated with simple VMD
form factors, we obtain $a_{\mu}^{\mathrm{LbyL;PS}} = (99 \pm 16) \times
10^{-11}$ for the sum of all light pseudoscalars. These values for the pion
and all pseudoscalars are about 20\% larger than the estimates obtained in
Refs.~\cite{BPP95, HKS95, HK98} which used other hadronic models for the form
factor.

As mentioned earlier, the identification of individual contributions, like
pion exchange, in hadronic light-by-light scattering is model-dependent as
soon as one uses off-shell form factors. We view our evaluation as being a
part of a full calculation based on a resonance Lagrangian, which fulfills all
the relevant QCD short-distance constraints, along the lines of the resonance
chiral theory approach developed in Ref.~\cite{Ecker_etal}.

We would like to stress that although our result for the pion-exchange
contribution is not too far from the value
$a_\mu^{\mathrm{LbyL};\pi^{0}-\mathrm{pole}} = (76.5 \pm 6.7) \times 10^{-11}$
given in Ref.~\cite{MV03},\footnote{Actually, using the on-shell LMD+V form
factor at the internal vertex with $h_2 = -10~\mbox{GeV}^2$ and $h_5 =
6.93~\mbox{GeV}^4$ and a constant (WZW) form factor at the external vertex, we
obtain $79.8 \times 10^{-11}$, close to the value $79.6 \times 10^{-11}$ given
in Ref.~\cite{BP07} and $79.7 \times 10^{-11}$ in
Ref.~\cite{Dorokhov_Broniowski}.} this is {\it pure coincidence}. We have used
off-shell LMD+V form factors at both vertices, whereas the authors of
Ref.~\cite{MV03} evaluated the {\it pion-pole} contribution using the on-shell
LMD+V form factor at the internal vertex and a constant (WZW) form factor at
the external vertex. On the other hand, as has been observed in
Refs.~\cite{BPP95, HKS95, HK98, KN01, BP07}, it is the region of momenta below
about 2~GeV which gives the bulk of the contribution to the final result in
the pion-exchange or pion-pole correction to hadronic light-by-light
scattering. This is also clearly visible from the weight functions that
multiply the form factors in the integrals and which have been presented in
Ref.~\cite{KN01}; see also Ref.~\cite{BP07}. They have a peak around
0.5~GeV. Therefore, as long as the absolute values of the model parameters
$h_3, h_4$ and $h_6$, which control the off-shellness of the pion in the LMD+V
form factor in Eq.~(\ref{KNpipioff}),\footnote{Note, however, that even if
$h_3, h_4$ and $h_6$ are put to zero, which actually violates the
short-distance constraint from Eq.~(\ref{constraint_h1_h3_h4}), one does not
recover the on-shell LMD+V form factor because of a term proportional to
$q_1^2 q_2^2 (q_1 + q_2)^2$ in the numerator in Eq.~(\ref{KNpipioff}).} are
not too large, i.e.\ below about 10 in appropriate units of GeV, one obtains a
result which will not be too far from the one obtained with on-shell LMD+V
form factors.  We have given some arguments for our choice of the parameters
$h_i$ and the ranges in which we vary them and they fulfill this constraint on
their size.  Recall that the constant term in the numerator of the form factor
in Eq.~(\ref{KNpipioff}) has the value $h_7 \simeq -14.8~\mbox{GeV}^6$.  As
pointed out before, our Ansatz for the neutral pion contribution to hadronic
light-by-light scattering with two off-shell LMD+V form factors agrees
qualitatively with the short-distance behavior derived in
Ref.~\cite{MV03}. However, since only the pion-pole contribution was
considered throughout that paper, a direct quantitative comparison is not
possible.

Recently, an evaluation of the pion-exchange contribution
appeared~\cite{Dorokhov_Broniowski} which uses a nonlocal chiral quark model
for the off-shell form factor ${\cal F}_{{\pi^0}^*\gamma^*\gamma^*}$. In that
model, off-shell effects of the pion always lead to a rather strong damping in
the form factor and the result $a_{\mu}^{\mathrm{LbyL};\pi^0} = (65 \pm 2)
\times 10^{-11}$ is therefore smaller than the pion-pole contribution obtained
in Ref.~\cite{MV03}. Although we also get a value which is (slightly) smaller
than the pion-pole contribution, our result depends on the chosen model
parameters, i.e.\ the constants $h_i$ in the LMD+V Ansatz and on the value for
the magnetic susceptibility $\chi(\mu)$.  We have given arguments for our
preferred choice of these parameters. In some other corner of the parameter
space one can, however, obtain a result which is larger than the pion-pole
contribution, i.e.\ we get a maximal value of $a_\mu^{\mathrm{LbyL};\pi^0} =
83.3 \times 10^{-11}$ in the scanned range.  Of course, any additional
information to pin down these model parameters and the ``correct'' value of
$\chi(\mu)$ and the relevant scale $\mu$ would be highly welcome to obtain a
more precise prediction for the pion-exchange contribution. At this point it
is not clear whether the nonlocal chiral quark model used in
Ref.~\cite{Dorokhov_Broniowski} or our LMD+V model for the form factor better
represents the strongly interacting region of QCD below about 2 GeV. At least
the LMD+V form factor fulfills all the relevant QCD short-distance
constraints. In any case, we think that the error of $\pm 2 \times 10^{-11}$
given in Ref.~\cite{Dorokhov_Broniowski} probably underestimates the inherent
uncertainly of any hadronic model.

In Ref.~\cite{MV03} an improved evaluation of the axial-vector contribution to
hadronic light-by-light scattering was given compared to
Refs.~\cite{BPP95,HKS95,HK98}, with the result $(22 \pm 5) \times
10^{-11}$. Note, however, that this seems to be again only the pole
contribution. Furthermore, Ref.~\cite{BPP95} obtained the following results
for the remaining contributions: $(-7 \pm 2) \times 10^{-11}$ for scalar
exchanges, $(-19 \pm 13) \times 10^{-11}$ for the dressed pion and kaon loops
and $(21 \pm 3) \times 10^{-11}$ for the dressed quark loops. These estimates
have more conservative errors than those given in
Refs.~\cite{HKS95,HK98}. Furthermore, the scalar-exchange contribution is not
evaluated in the latter references. If we combine our value for the
pseudoscalars with these results, we obtain the new estimate
\be
a_{\mu}^{\mathrm{LbyL; had}} = (116 \pm 40) \times 10^{-11} 
\ee
for the total hadronic light-by-light scattering contribution to the anomalous
magnetic moment of the muon. To be conservative, we have added all the errors
linearly, as has become customary in recent years, and rounded up the obtained
value $\pm 39 \times 10^{-11}$. In the very recent
review~\cite{Prades_deRafael_Vainshtein09} the central values of some of the
individual contributions to hadronic light-by-light scattering are adjusted
and some errors are enlarged to cover the results obtained by various groups
which used different models. The errors are finally added in quadrature to
yield the estimate $a_{\mu}^{\mathrm{LbyL; had}} = (105 \pm 26) \times
10^{-11}$. Note that the dressed light quark loops are not included as a
separate contribution in Ref.~\cite{Prades_deRafael_Vainshtein09}. They are 
assumed to be already covered by using the short-distance constraint from
Ref.~\cite{MV03} on the pseudoscalar-pole contribution. 

Some progress has been achieved in recent years to better understand the
hadronic light-by-light scattering contribution to the muon $g-2$. We hope
that our new short-distance constraint on the off-shell form factor ${\cal
F}_{{\pi^0}^*\gamma^*\gamma}(q^2, q^2, 0)$ at the external vertex will further
help to control the numerically dominant pion-exchange contribution.  We
should not forget, however, that the contribution of the exchanges of $\eta$
and $\eta^\prime$ are theoretically not that well understood. We have
simply used VMD form factors as has been done in most other evaluations. A new
analysis, along the lines of the approach for the pion, is definitely
needed. Finally, as stressed in Refs.~\cite{BP07,Prades_deRafael_Vainshtein09}
a better control of the numerically subdominant but non-negligible other
contributions is also needed, if we fully want to profit from a potential
future $g-2$ experiment.

\section*{Acknowledgments} 

I am grateful to F.\ Jegerlehner for pointing out that fully off-shell form
factors should be used to evaluate the pion-exchange contribution, for helpful
discussions and for numerous correspondences. Furthermore, I would like to
thank G.\ Colangelo, J.\ Gasser, M.\ Knecht, H.\ Leutwyler, P.\ Minkowski, M.\
Perrottet, E.\ de Rafael and A.\ Vainshtein for stimulating discussions. I am
grateful to the Institute for Theoretical Physics at the University of Bern,
the Centre de Physique Th\'eorique in Marseille, the Theory Group at DESY
Zeuthen and the Institute for Theoretical Physics at ETH Z\"urich for the warm
hospitality during some stages of working on this project. This work was
supported by the Department of Atomic Energy, Government of India, under a
5-Years Plan Project.

\section*{Appendix}
\renewcommand{\theequation}{A\arabic{equation}}
\setcounter{equation}{0}

We provide in this appendix a parametrization of $a_\mu^{\mathrm{LbyL};\pi^0}$
for arbitrary coefficients $h_i$ in the LMD+V Ansatz for the off-shell form
factor ${\cal F}_{{\pi^0}^*\gamma^*\gamma^*}$ in Eq.~(\ref{KNpipioff}). This
will facilitate future updates of our result for the pion-exchange
contribution in Eq.~(\ref{amupi0LMD+V}) in case some of the parameters $h_i$
or the value (and the relevant scale) of the magnetic susceptibility
$\chi(\mu)$ will be known more precisely.

Measuring the parameters $h_i$ in appropriate units of GeV, i.e.\ defining
$\tilde h_i = h_i / \gev^2$ for $i=1,2,3,4$, $\tilde h_i = h_i /
\gev^4$ for $i=5,6$ and $\tilde h_7 = h_7 / \gev^6$, we can write
\be \label{general_param}
a_\mu^{\mathrm{LbyL};\pi^0} = \left( {\alpha \over \pi} \right)^3 \, \left[
\sum_{i=1}^{7} c_i \, \tilde h_i 
+ \sum_{i=1}^{7} \sum_{j=i}^{7} c_{ij} \, \tilde h_i \, \tilde h_j  
  \right], 
\ee
where the coefficients $c_i$ and $c_{ij}$ are given in
Table~\ref{tab:ci_cij}. 
\begin{table}[h] 
\caption{Values of the coefficients $c_i$ and $c_{ij}$ which appear in
  Eq.~(\ref{general_param}).}   
\begin{center} 
\renewcommand{\arraystretch}{1.25}
\begin{tabular}{|c|r@{.}lr@{.}lr@{.}lr@{.}lr@{.}lr@{.}lr@{.}l|}
\hline  
& \multicolumn{2}{c}{$i=1$} & \multicolumn{2}{c}{$i=2$} &
\multicolumn{2}{c}{$i=3$} & \multicolumn{2}{c}{$i=4$} &
\multicolumn{2}{c}{$i=5$} & \multicolumn{2}{c}{$i=6$} &
\multicolumn{2}{c|}{$i=7$} \\ 
\hline 
$c_i \times 10^4$
& $-1$ & 4530 & \multicolumn{2}{c}{0} & $-1$ & 4530 & $-1$ & 4530 & 0 & 4547 & 
0 & 4547 & $-1$ & 2048 \\ 
\hline
\hline 
$c_{ij} \times 10^4$ & 
\multicolumn{2}{c}{$j=1$} & \multicolumn{2}{c}{$j=2$} &
\multicolumn{2}{c}{$j=3$} & \multicolumn{2}{c}{$j=4$} &
\multicolumn{2}{c}{$j=5$} & \multicolumn{2}{c}{$j=6$} &
\multicolumn{2}{c|}{$j=7$} \\ 
\hline 
$i=1$ & 0 & 4447 & 0 & 0729 & 0 & 7428 & 0 & 7120 & $-0$ & 3620 & $-0$ & 3332
& 0 & 9916 \\  
$i=2$ & \multicolumn{2}{c}{$\cdots$}     & \multicolumn{2}{c}{0}      & 0 &
0730 & 
0 & 0729 & $-0$ & 0557 & $-0$ & 0557 & 0 & 2221 \\ 
$i=3$ & \multicolumn{2}{c}{$\cdots$}     & \multicolumn{2}{c}{$\cdots$}     &
0 & 2980 & 0 & 5653 & $-0$ & 1967 & $-0$ & 1679 & 0 & 1796 \\ 
$i=4$ & \multicolumn{2}{c}{$\cdots$}     & \multicolumn{2}{c}{$\cdots$}     &
\multicolumn{2}{c}{$\cdots$}     & 0 & 2672 & $-0$ & 1679 & $-0$ & 1391 & 0 &
1162 
\\  
$i=5$ & \multicolumn{2}{c}{$\cdots$}     & \multicolumn{2}{c}{$\cdots$}     &
\multicolumn{2}{c}{$\cdots$}     & \multicolumn{2}{c}{$\cdots$}     & 0 & 1215
& 0 & 1796 & $-0$ & 8072 \\ 
$i=6$ & \multicolumn{2}{c}{$\cdots$}     & \multicolumn{2}{c}{$\cdots$}     &
\multicolumn{2}{c}{$\cdots$}     & \multicolumn{2}{c}{$\cdots$}     &
\multicolumn{2}{c}{$\cdots$}      & 0 & 0581  & $-0$ & 3052 \\ 
$i=7$ & \multicolumn{2}{c}{$\cdots$}     & \multicolumn{2}{c}{$\cdots$}     &
\multicolumn{2}{c}{$\cdots$}     & \multicolumn{2}{c}{$\cdots$}     &
\multicolumn{2}{c}{$\cdots$}     & \multicolumn{2}{c}{$\cdots$}      & 1 &
6122 \\  
\hline 
\end{tabular}
\label{tab:ci_cij}
\end{center} 
\end{table}

This representation follows immediately from the general expression for the
pion-exchange contribution in Eq.~(\ref{a_pion_2}) and from the LMD+V Ansatz
for the form factor in Eq.~(\ref{KNpipioff}). Because of our new
short-distance constraint at the external vertex, the parameters $h_1, h_3$
and $h_4$ are not independent, but must obey the relation $h_1 + h_3 + h_4 =
M_{V_1}^2 \, M_{V_2}^2 \, \chi$; see Eq.~(\ref{constraint_h1_h3_h4}). Note the
absence of a term without the constants $h_i$ in
Eq.~(\ref{general_param}). This follows from the fact that at the external
vertex with the soft photon the form factor ${\cal
F}_{{\pi^0}^*\gamma^*\gamma}(q_3^2, q_3^2,0)$ enters, e.g.\ in the term with
$T_2$ in Eq.~(\ref{a_pion_2}). This also leads to $c_2 = 0$ and $c_{22} = 0$
in Table~\ref{tab:ci_cij}.  For evaluating the integrals, we used $m_\mu =
105.658369~\mev$, $m_{\pi^0} = 134.9766~\mev$, $F_\pi = 92.4~\mev$, $M_{V_1} =
M_\rho = 775.49~\mev$ and $M_{V_2} = M_{\rho^\prime} = 1.465~\gev$. The decay
constant $F_\pi$ only enters as an overall factor in ${\cal
F}_{{\pi^0}^*\gamma^*\gamma^*}$.

Note, however, that some of the model parameters $h_i$ are quite well
determined from experimental data and theoretical constraints.  The
normalization with the pion decay amplitude $\pi^0 \to \gamma\gamma$ yields
$h_7 = - N_C M_{V_1}^4 M_{V_2}^4 / (4 \pi^2 F_\pi^2) - h_6 m_\pi^2 - h_4
m_\pi^4$. The Brodsky-Lepage behavior can be reproduced by choosing $h_1 =
0~\mbox{GeV}^2$. Furthermore, a fit to the CLEO data for the on-shell form
factor ${\cal F}_{{\pi^0}\gamma^*\gamma}^{\rm LMD+V}(m_\pi^2, -Q^2, 0)$ leads
to $h_5 = (6.93 \pm 0.26)~\mbox{GeV}^4 - h_3 m_\pi^2$. Finally, the constant
$h_2$ can be obtained from higher-twist corrections in the OPE, with the
result $h_2=-4\,(M_{V_1}^2+M_{V_2}^2) + (16/9)\,\delta^2 \simeq
-10.63~\gev^2$.

If we use the above informations to fix $h_1, h_2, h_5$ and $h_7$, we obtain
the simplified expression
\bea
a_\mu^{\mathrm{LbyL};\pi^0} & = & \left( {\alpha \over \pi} \right)^3 
\left[  
\ 503.3764 - 6.5223 \, \tilde h_3 - 5.0962 \, \tilde h_4 + 7.8557
\, \tilde h_6  \right. \nonumber \\
& & \hspace*{1.4cm} 
+ 0.3017 \, \tilde h_3^2 + 0.5683 \, \tilde h_3 \, \tilde h_4 - 0.1747 \,
\tilde h_3 \, \tilde h_6 \nonumber 
\\[0.1cm] 
& & \hspace*{1.35cm} 
\left. + 0.2672 \, \tilde h_4^2 - 0.1411 \, \tilde h_4 \, \tilde h_6 
+ 0.0642 \, \tilde h_6^2 \  
\right] \times 10^{-4}, 
\eea
where only $h_3, h_4$ and $h_6$ enter as free parameters. Note again, however,
that $h_3$ and $h_4$ are not independent, but now obey the relation $h_3 + h_4
= M_{V_1}^2 \, M_{V_2}^2 \, \chi$.

\end{document}